\documentclass[aps,pre,twocolumn,groupedaddress,amsmath,amssymb,superscriptaddress,reprint,showpacs,showkeys,nofootinbib]{revtex4-1}

\usepackage{graphicx}
\usepackage{dcolumn}
\usepackage{color}
\usepackage{epsfig}
\usepackage{adjustbox}
\usepackage{subfigure}
\usepackage{amsmath}
\usepackage{multirow}
\usepackage{bm}
\usepackage{amsthm}
\usepackage{amsfonts}
\usepackage{float}
\usepackage{times}
\usepackage{amssymb}
\usepackage{relsize}
\usepackage{booktabs}
\usepackage[normalem]{ulem}
 
\usepackage[colorlinks=true,citecolor=blue,linkcolor=blue,urlcolor=blue]{hyperref}

\usepackage[sort&compress]{natbib}

\begin{document}
	
\title{Quantum Brain Networks: a Perspective}

\author{E. R. Miranda}
\affiliation{Interdisciplinary Centre for Computer Music Research (ICCMR), University of Plymouth, United Kingdom}

\author{S. Venkatesh}
\affiliation{Interdisciplinary Centre for Computer Music Research (ICCMR), University of Plymouth, United Kingdom}

\author{C. Hernani-Morales}
\affiliation{IDAL, Electronic Engineering Department, University of Valencia,
Avgda. Universitat s/n, 46100 Burjassot, Valencia, Spain}

\author{L. Lamata}
\affiliation{Departamento de F\'isica At\'omica, Molecular y Nuclear, Universidad de Sevilla, 41080 Sevilla, Spain}

\author{J. D. Mart\'in-Guerrero}
\affiliation{IDAL, Electronic Engineering Department, University of Valencia,
Avgda. Universitat s/n, 46100 Burjassot, Valencia, Spain}

\author{E. Solano}
\email[Email:]{\qquad enr.solano@gmail.com}
\affiliation{Department of Physical Chemistry, University of the Basque Country UPV/EHU, Apartado 644, 48080 Bilbao, Spain}
\affiliation{IKERBASQUE, Basque Foundation for Science, Plaza Euskadi 5, 48009 Bilbao, Spain}
\affiliation{International Center of Quantum Artificial Intelligence for Science and Technology (QuArtist) \\ and Physics Department, Shanghai University, 200444 Shanghai, China}
\affiliation{Kipu Quantum, Kurwenalstrasse 1, 80804 Munich, Germany}

\date{\today}

\begin{abstract}

We propose Quantum Brain Networks (QBraiNs) as a new interdisciplinary field integrating knowledge and methods from neurotechnology, artificial intelligence, and quantum computing. The objective is to develop an enhanced connectivity between the human brain and quantum computers for a variety of disruptive applications. We foresee the emergence of hybrid classical-quantum networks of wetware and hardware nodes, mediated by machine learning techniques and brain-machine interfaces. QBraiNs will harness and transform in unprecedented ways arts, science, technologies, and entrepreneurship, in particular activities related to medicine, Internet of humans, intelligent devices, sensorial experience, gaming, Internet of things, crypto trading, and business. 

\end{abstract}

\maketitle

\section{Introduction}
\label{sec:Intro}

Since the origin of quantum physics, the role of human observers in the disturbing collapse of the wave function occupies a central role. Challenges to our classical intuition led to a series of proposed paradoxes, mostly due to the extrapolation of microscopic quantum phenomena to our distinct macroscopic human experience. Counterintuitive Gedankenexperiments, as the celebrated cases of the Schr\"odinger's Cat~\cite{Schrodinger1935} and Wigner's Friend~\cite{Wigner1967}, illustrate the historical difficulties in assuming the consequences of the quantum theory~\cite{FrauchigerRenner}. Moreover, risky conjectures have also been raised concerning possible quantum phenomena in brain processes, in particular for making sense of human free will, mind models, decision-making, and consciousness~\cite{Penrose1994,Fisher2015,Jedlicka2017}. In this sense, it would be scientifically and technologically groundbreaking, from fundamentals of hardware and wetware science to cutting-edge applications, the development of a closer connection between human brains and quantum computers (QCs). However, our understanding of the brain, mind, and whatever we may mean by consciousness, is still rudimentary. This makes it difficult to interfacing the brain directly with external quantum devices or quantum processors~\cite{Feynman1982,Nielsen2010}. Nevertheless, artificial intelligence (AI) may come at our rescue to achieve this otherwise impossible task at this point of the 21st century.

In the last decades, we may find bottom-up approaches to consider the merge of biological properties with quantum phenomena. In the case of quantum biology, possible quantum features might explain efficiency of photosynthesis~\cite{QuantumBiology}. Also, neuromorphic technologies are being studied to save energy and enhance AI applications~\cite{NeuromorphicComputing}. More recently, bio-inspired quantum artificial life has been proposed and implemented in a quantum computer~\cite{QAL}, while neuromorphic quantum computing using quantum memristors for quantum neural networks is getting first experimental tests~\cite{QuantumMemristors2016,AtomicQuantumMemristor2021,PhotonicQuantumMemristor2021}.

\begin{figure}[t!]
	\includegraphics[width=0.9\linewidth]{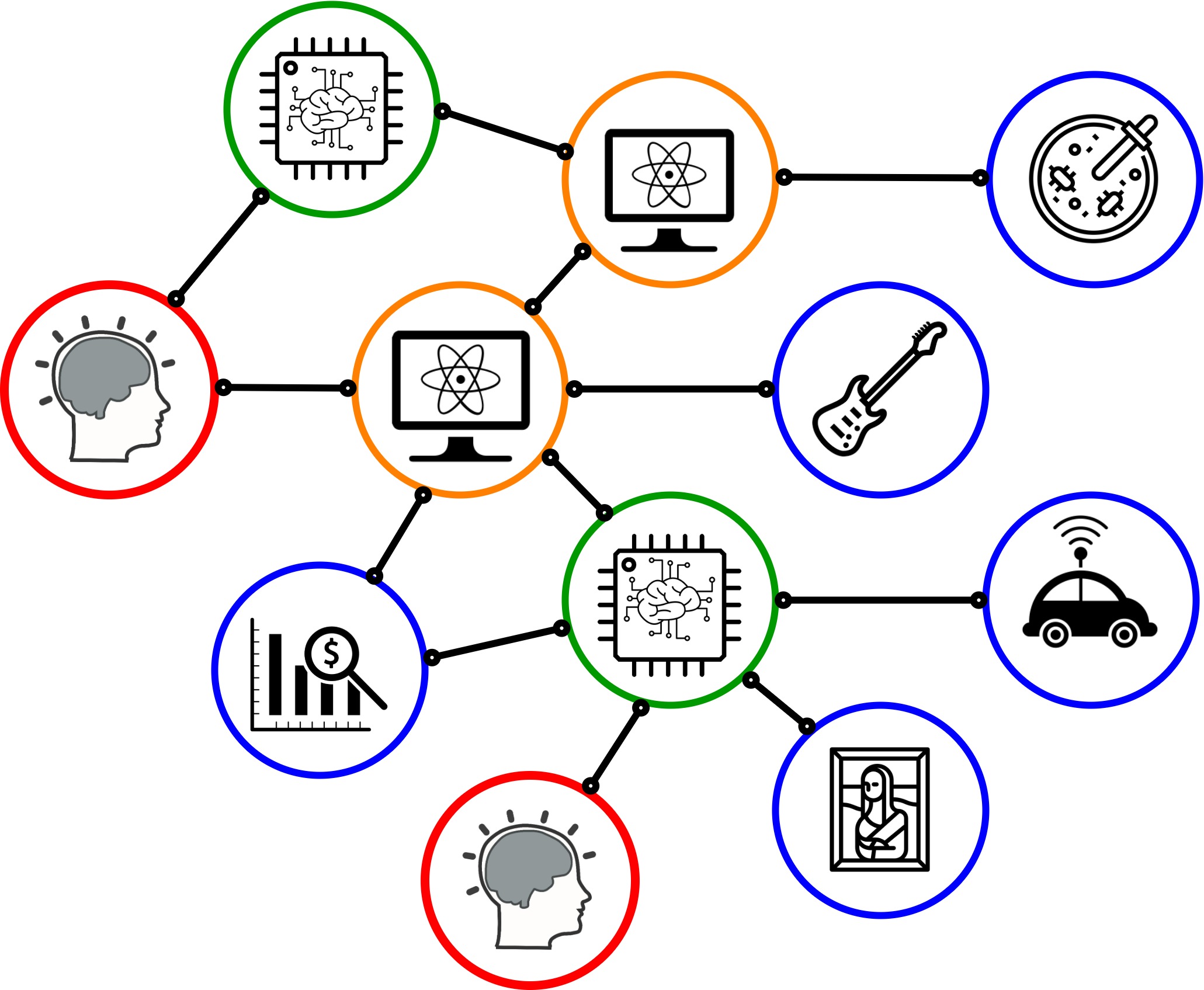}
	\caption{Quantum Brain Networks}
	\label{fig:fig1}
\end{figure}

In this Perspective, we explore and argue in favor of the technical feasibility of a novel interdisciplinary field we coined as Quantum Brain Networks (QBraiNs), a top-down approach based on solid ground and innovative merge of neurotechnology, artificial intelligence, and quantum computers.

{\it Definition:} a QBraiN is a multidirectional wired/wireless network of local/distributed nodes with wetware/hardware classical/quantum processors.

We take it for granted that such a definition of a QBraiN will need suitable adaptations when implemented in different technological platforms and applied to different use cases. We also assume several parts of the same human brain may be considered as QBraiN nodes. It is noteworthy to mention that every step in the development of QBraiN models or building prototypes may settle decisive milestones in challenging domains. Among them, we could mention philosophy of science, fundamentals of physics, brain science, mind and consciousness models, augmented classical-quantum sensorial experience, enhanced classical-quantum computing, and innovative wetware-hardware applications (Fig. ~\ref{fig:fig1}).

\section{Neurotechnology}
\label{sec:NTs}

Neurotechnology enables us to understand the human brain and mind, while improving and repairing their accessible functions. For the sake of simplicity, we associate here brain to hardware and mind to software. Along these lines, brain scanning technologies, such as EEG~\footnote{EEG stands for electroencephalogram. It refers to the brain's spontaneous electrical activity over a period of time, as recorded from multiple electrodes placed on the scalp or inside the skull.} and fMRI~\footnote{fMRI stands for functional Magnetic Resonance Imaging. It measures brain activity by detecting changes associated with blood flow.}, allow clinicians and researchers to visualise the brain and monitor its activity. Increasingly sophisticated brain scanning technologies are also enabling the development of innovative brain-machine interfaces (BMIs~\footnote{Also referred to as BCIs or Brain-Computer Interfaces.}). These devices endow humans with the ability to control machines by thinking, detectable as brain activity, without the physical constraints of the body~\cite{IntroBCI}. The ultimate goal of BMI research is to network brains and machines to directly communicate with each another. Despite tremendous research progress and encouraging demonstrations of possible applications, from typewriting texts~\cite{SpellingBCI} to making music~\cite{EduardoBook}, BMI systems are not widely available for general use yet. In fact, currently, brain-machine communication takes place mostly from humans to machines, leaving still open bidirectional and more integrated networks for further developments. In this sense, it is noteworthy to mention that machines can only communicate with us through our senses. The development of robust connectivity of our nervous system with devices outside our body is hampered by the lack of adequate transducers, and the inability to electronically compute humongous amount of data in real time. Emerging ultra-sensitive quantum brain sensors~\cite{QuantumBrainSensor} and quantum computers with processing speeds beyond state-of-the-art classical supercomputers are bound to impact significantly on BMI research~\cite{EduardoMasterpiece}. The success of quantum technologies will enable the development of new hybrid classical-quantum networks of wetware and hardware, enhancing intelligent creativity and communication.

\section{Artificial Intelligence}
\label{sec:AI}

Since the origin of humankind, the use of tools is ubiquitous in all human activities, be for survival, leisure, arts, or technologies. These tools have reached sophisticated levels of abstraction with the emergence of intelligent devices. They cannot only perform a given task but also communicate with other machines in order to solve challenging problems~\cite{iot}.  The concept of AI as machines that can perform smart tasks appears naturally in this framework.  These machines and intelligent devices are the source of an ever-increasing amount of information,  that oftentimes is given by means of computable data. This implies the use of data that can be analysed by machines in order to extract information leading to knowledge discovery,  that is Machine Learning (ML)~\cite{alpaydin_ML}.

AI,  and particularly ML, has grown steadily in the last decade. Its success is mostly due to the fact that Deep Learning (DL)~\cite{DL_nature}, a ML approach based on models that may be characterized by millions of parameters, has shown its ability to solve cognitive tasks that were computationally intractable until very recently. Among them, we could mention image segmentation and classification, natural language processing, sentiment analysis, generative art and music~\footnote{Here we refer to computer-generated pieces of art and music of equivalent quality of human-made ones.}, and autonomous vehicles.  One may wonder how the solutions to these problems may be harnessed further by means of networking with quantum computing technologies.

The bottleneck of AI progress has been frequently associated with hardware development.  The situation is different at present, where ML innovations strongly depend on the involved creativity to tackle a given problem.  Furthermore, an important enhancement to AI may happen if we are able to involve features of neurotechnology and, even more, if we may borrow key concepts from quantum computing.

\section{Quantum Computers}
\label{sec:QCs}

The experimental success of quantum physics, despite difficulties for an intuitive interpretation, led to consider quantum computers as an innovative paradigm for efficiently simulating and computing the distinct quantum aspects of nature~\cite{Feynman1982}. Two decades ago, quantum information led the transition from single-particle tests towards multiqubit experiments, then came the emergence of quantum computers in a variety of platforms, while last years have brought ramping worldwide investments in quantum technologies. Recently, the first claims have been made on quantum supremacy~\cite{Martinis2019,Pan2020}, that is quantum processors outperforming classical computers without a request for useful applications yet. Nowadays, there is an agitated debate on whether reaching quantum advantage, quantum processors outperforming classical computers for industrial cases, might be possible in the near future. If achieved, currently impossible solutions for applications such as molecular design, financial modeling, optimization problems, and machine learning may be at reach~\cite{Nori2014,Biamonte2017,Aspuru-Guzik2020}. 

Some of the quantum platforms that are currently being considered are superconducting circuits, trapped ions, cold atoms, photonics, quantum dots, spin impurities in solids, nitrogen-vacancy centers in diamond, as well as nuclear magnetic resonance. Each of them has advantages and disadvantages depending on their specific applications, and they have been evolving in parallel at a different speed in the past three decades. In this sense, any requirement for a possible control of a quantum computer with human brain activity would end up in a specific choice of technology. Fortunately, there are also intense activities at the educational, professional, and business level, for developing web-based classical simulators of remote quantum computers.

\section{Quantum Brain Networks}
\label{sec:QBNs}

We consider the current availability and reachability of the theoretical methods and technological requirements for designing, building, and using QBraiNs for a variety of interdisciplinary applications. Accordingly, we provide below three subsections containing our vision on QBraiNs, their unique connection possibilities for a variety of situations, and potentially disruptive use cases.

\subsection{\it Vision on QBraiNs}
\label{subsec:VisionQBNs}

We propose a list of logical statements with the aim of arguing in favor of our final position on QBraiNs.

\noindent
1) {\it It is possible to read and harness brain data with BMIs}

\noindent
BMIs can be accessed commercially with an increasing number of channels, with different kinds of technologies and levels of complexity. For instance, there are commercially available portable EEG devices ranging from simple systems for gaming to sophisticated designs for research and medical purposes. There are invasive or non-invasive BMI systems, as well as research to develop wireless devices~\cite{WirelessBCI2021}.

\noindent
2) {\it Brain data can be processed by classical computers via ML}

\noindent
The communication between brains and machines, be digital or analog, is a historic conundrum. Essentially, this problem is due to the lack of trustful and minimally complex brain and mind models, as well as inadequate transducers. AI brings pragmatic solutions to this mismatch, not by proposing a structured language but through bypassing it. There is rudimentary BMI technology available today to build simple brain-controlled systems using ML running on classical computers, including laptops, tablets, and mobile phones~\footnote{An example: \url{https://www.volvocars.com/uk/about/humanmade/discover-volvo/music-of-the-mind}}.

\noindent
3) {\it Classical computers can control quantum computers}

\noindent
There is an increasing number of open-source software platforms provided by quantum computing companies and startups to control quantum computer simulators and real quantum computers via the World Wide Web (see, as paradigmatic examples, Qiskit from IBM and Cirq from Google). The worldwide success  and spread of such web-based services has created a revolution in the dissemination of possible quantum computing solutions for important scientific, industrial, and societal applications.

\noindent
4) {\it Human brains can be connected to quantum computers~\footnote{Experimental work and a proof-of-concept are in progress by the authors of this Perspective.}}

\noindent
As a natural corollary to the previous items, we propose that human brains can be directly connected to quantum computers with currently available concepts, methods, and technologies. In this manner, QBraiNs are bound to become a reality and a playground for novel arts, leisure, science, technology, and entrepreneurial activities.

\subsection{\it Connectivity of QBraiNs}
\label{subsec:ConnectivityQBNs}

Depending on specific goals, different possibilities to connect classical and quantum nodes of a QBraiN may emerge.

\noindent
1) {\it One-to-one bidirectional brain-QC transducer}

\noindent
A basic QBraiN will involve a single human brain, possibly with several BMI nodes, connected to a quantum computer. The suitable choice of technology may be decided on the basis of latency, signal processing, and selected use case.

\noindent
2) {\it Multiple brain-QC nodes}

\noindent
Local and distributed QBraiN nodes will provide a wetware-hardware classical-quantum processing network with unprecedented computational power and capabilities. 

\noindent
3) {\it Hybrid brain-SC-QC}

\noindent
A natural extension will be to add to the QBraiN one or several nodes involving supercomputing and high-performance cloud-based facilities.

\subsection{\it Applications of QBraiNs}
\label{subsec:ApplicationsQBNs}

Here we present a first tentative list of applications of QBraiNs, mostly related to some key forefronts in science and technology.

\noindent
1) {\it Fundamental brain-QC tests}

\noindent
QBraiNs will allow interdisciplinary researchers and engineers to realize fundamental tests related to quantum physics, quantum information, psychology and behavioural models, quantum measurement and human perception in general~\cite{BCIbehavior2021}. 

\noindent
2) {\it Brains controlling hybrid classical-quantum computers}

\noindent
Distributed nodes of wetware-hardware classical-quantum processors will allow QBraiN developers and practitioners to explore enhanced computational power and communication in Internet of humans, Internet of things, and the crypto space.

\noindent
3) {\it Quantum-enhanced brains}

\noindent
QBraiNs will allow humans to enter into a new era of interconnectivity with machines. We envision a wide class of possibilities for enhancing analysis and creativity, as well as intuitive and rational control of intelligent classical-quantum systems.

\noindent
4) {\it Quantum Arts}

\noindent
QBraiNs will allow artists and spectators a unique two-level sensorial quantum experience. A first quantum level will be related to the signal sequence: classical input, quantum filtering, classical output. And a a second quantum level will involve quantum feedback of looped emotions and thoughts.

\section{Concluding Remarks}
\label{sec:Conclusions}

We have discussed the conceptual and technological readiness for a promising interdisciplinary field, Quantum Brain Networks (QBraiNs), involving the hybrid classical-quantum communication between human brains, intelligent machines, and quantum computers. Beyond the necessary upgrades in upcoming hardware and wetware developments, including transducers and control systems, we believe QBraiNs will have a wide variety of disruptive applications in association with versatile connectivity options. In this sense, we have provided feasible use cases where wetware and hardware interconnectivity, at the classical and quantum levels, would be transformational for current processing and communication technologies. This Perspective should open wide discussions in yet unexplored directions of the long-term relation between humans and machines. Finally, we understand that QBraiNs may open another path of ethical issues to consider, an important topic that goes beyond the scope of this Perspective.

\section{Acknowledgments}

We acknowledge support from QMiCS (820505) and OpenSuperQ (820363) projects of the EU Flagship on Quantum Technologies, National Natural Science Foundation of China (NSFC) (12075145), STCSM (2019SHZDZX01-ZX04, 18010500400 and 18ZR1415500), Spanish Government PGC2018-095113-B-I00, PID2019-104002GB-C21 and PID2019-104002GB-C22 (MCIU/AEI/FEDER, UE), Basque Government IT986-16, EU FET Open Grant Quromorphic and EPIQUS. We thank the ICCMR, University of Plymouth, UK, for support and access to BMI resources.

\end{document}